\input harvmac
\input psfig

%
%
%

\def\tilde{\widetilde}
\def\bar{\overline}
\def\hat{\widehat}
\def\*{\star}
\def\[{\left[}
\def\]{\right]}
\def\({\left(}		
\def\){\right)}

%
%

\def\frac#1#2{{#1 \over #2}}
\def\inv#1{{1 \over #1}}

\def\d{\partial}

\def\2pi{\hbox{$2\pi i$}}

\def\dsl{\raise.15ex\hbox{/}\kern-.57em\partial}
\def\Dsl{\,\raise.15ex\hbox{/}\mkern-.13.5mu D}
%
%

\def\ep{\epsilon}

%
%

		\def\CR{{\cal R}}
\def\CS{{\cal S}}

\def\2pi{\hbox{$2\pi i$}}

\def\dsl{\raise.15ex\hbox{/}\kern-.57em\partial}
\def\Dsl{\,\raise.15ex\hbox{/}\mkern-.13.5mu D}
%
%
%
\font\numbers=cmss12
\font\upright=cmu10 scaled\magstep1
\def\stroke{\vrule height8pt width0.4pt depth-0.1pt}
\def\topfleck{\vrule height8pt width0.5pt depth-5.9pt}
\def\botfleck{\vrule height2pt width0.5pt depth0.1pt}
\def\Zmath{\vcenter{\hbox{\numbers\rlap{\rlap{Z}\kern
0.8pt\topfleck}\kern
2.2pt
                   \rlap Z\kern 6pt\botfleck\kern 1pt}}}
\def\Qmath{\vcenter{\hbox{\upright\rlap{\rlap{Q}\kern
                   3.8pt\stroke}\phantom{Q}}}}
\def\Nmath{\vcenter{\hbox{\upright\rlap{I}\kern 1.7pt N}}}
\def\Cmath{\vcenter{\hbox{\upright\rlap{\rlap{C}\kern
                   3.8pt\stroke}\phantom{C}}}}
\def\Rmath{\vcenter{\hbox{\upright\rlap{I}\kern 1.7pt R}}}
\def\Z{\ifmmode\Zmath\else$\Zmath$\fi}
\def\Q{\ifmmode\Qmath\else$\Qmath$\fi}
\def\N{\ifmmode\Nmath\else$\Nmath$\fi}
\def\C{\ifmmode\Cmath\else$\Cmath$\fi}
\def\R{\ifmmode\Rmath\else$\Rmath$\fi}

\Title{CLNS 96/1442; hep-th/9701016}
{\vbox{\centerline{The Scattering Theory of}
\centerline{Oscillator Defects in an Optical Fiber}}}

\bigskip

\centerline{Robert Konik and Andr\'{e} LeClair}
\medskip\centerline{Newman Laboratory}
\centerline{Cornell University}
\centerline{Ithaca, NY  14853}
\bigskip\bigskip

\vskip .1in

We examine harmonic oscillator defects
coupled to a photon field in
the environs of an optical fiber.  Using techniques borrowed or extended
from the theory of two dimensional quantum fields with boundaries and
defects,
we are able to compute exactly a number of interesting
quantities.  We calculate the scattering S-matrices (i.e. the reflection
and transmission amplitudes) of the photons off a single defect.  We
determine using techniques derived from thermodynamic Bethe ansatz 
(TBA) the thermodynamic
potentials of the interacting photon-defect system.  And we compute
several correlators of physical interest.  We find the photon occupancy
at finite temperature, the spontaneous emission spectrum from the decay
of an excited state, and the correlation functions of the defect degrees
of freedom.  
In an extension of the single defect theory, we
find the photonic band structure that arises from a 
periodic array of harmonic oscillators.
In another
extension, we examine a continuous array of defects and exactly
derive its dispersion relation.  With some differences,
the spectrum is similar to
that found for EM wave propagation in covalent crystals.
We then add to this continuum theory isolated defects, so as to obtain a
more realistic model of defects embedded in a frequency dependent
dielectric medium.  We do this both with a single isolated defect and with
an array of isolated defects, and so compute how 
the S-matrices and the band structure change in a
dynamic medium.

\vskip .1in

\Date{12/96}

\noblackbox

\newsec{Introduction}

In this paper we examine photons interacting with harmonic
oscillator impurities in an optical fiber.  
Because
this problem is inherently two dimensional, we are able to attack
it by applying or extending
various techniques from two dimensional defect and boundary 
field theory (\ref\muss{G. Delfino, G. Mussardo, and P. Simonetti,
Nucl. Phys. B432 (1994) 518.} 
\ref\zam{S. Ghoshal and 
A. Zamolodchikov, Int. J. Mod. Phys. A9 (1994) 3841.}
\ref\tba{A. LeClair, G. Mussardo, H. Saleur, S. Skorik, 
Nucl. Phys. B453 (1995) 581.}).
This permits us to compute exactly quantities previously inaccessible.  
Here, for example, using methods akin to
thermodynamic Bethe ansatz (TBA), we compute 
thermodynamic potentials.  And through extending the notion of boundary
states, we exactly compute correlators.  

Normally the impurity atoms in optical problems are modeled as two-level
systems.  Although this makes the problem non-linear, some success has been
had in finding exact solutions in two dimensions.  Under the rotating wave
approximation, a continuum array of defects has been shown to be integrable
(\ref\rup{V. I. Rupasov, JETP Lett. 36 (1982) 142.}).  
There, using algebraic Bethe ansatz, the spectrum and the
S-matrices of the theory were found. In 
\ref\and{A. LeClair, F. Lesage, S. Lukyanov, H. Saleur, {\it
Exact Solution of the Maxwell-Bloch Theory in Quantum Optics}, CLNS
preprint.} it was shown that
a single two-level atom can be mapped onto a Kondo impurity.  With this
mapping they were able to compute thermodynamic potentials exactly and
to obtain form factor expansions of some correlators.  But such
successes are limited.  Because of the theory's non-linearity,
integrability is key to obtaining exact solutions.  However integrability
limits the type of defect configurations one can consider.  Outside the
rotating wave approximation, an array of two-level atoms is not known to be 
integrable.  To then say something one must resort to perturbation
theory (see \ref\per{A. LeClair, {\it QED for a Fibrillar Medium
of Two-Level Atoms}, hep-th 9604100.}).  And although a single two-level atom
can be treated, two or some larger number cannot.  Nor does integrability
guarantee that all interesting physical quantities can be accessed.  Photon
correlators for two-level systems are still only known approximately.  And
the computation of correlators involving
the impurity degrees of freedom remains an open
problem.  To sidestep these difficulties we choose to model the atoms as
harmonic oscillators.  In doing this we cease to be limited by particular
configurations of atoms under certain approximations and we gain access to
a wider variety of quantities, including both exact photon and impurity
correlators.

The replacement of two-level atoms by harmonic oscillators is not
necessarily a bad one.  Providing the intensity of the photon field in
the fiber is low, only the oscillator's ground and first excited state are
likely to be occupied.  Under these conditions, the oscillator mimics a
two-level system.  We will see later, for example, how the spontaneous 
emission spectrum mimics that of a two-level atom.  However, we cannot
hope to model such phenomena as strong field resonance fluorescence.

We exploit our ability to study arbitrary configurations of defects by
examining both periodic and continuum arrays of defects.  With a periodic
array of defects, a band structure of photons naturally appears.  
This band structure is equipped with gaps of two types: one arising from 
being on resonance and a more typical one of gaps
appearing at the edges of the Brillouin zones.

In the case of the continuum array of defects, we derive a dispersion
relation for the photons propagating among the defects.  We find that
the dressed photons are described by the typical polariton dispersion
relation.  Again exploiting our configurational
flexibility, we examine
theories where isolated defects are added to the continuum array.  We
look at two examples: an isolated defect and a periodic array of defects.
With the isolated defect we compute the scattering matrices of the polaritons
off the defect.  In the case of the periodic array, we calculate the band
structure obeyed by the polaritons.

The ability to consider defects living in a dynamic medium lays 
the ground for the 
consideration of photon localization.  By choosing the energy of an isolated 
defect to lie within the gap between the two polariton branches, there is
some chance to find a photon in a bound state.  We will consider this 
possibility in an upcoming work.

The paper is organized as follows.  In the first section an overview of the
model as well as its reduction to two dimensions is given.  In the next
section the scattering matrices are derived, followed by a section
giving the TBA analysis of the thermodynamic potentials.  In the next
section correlators are computed.  There we compute both time ordered and
non-time ordered correlators.  The non-time ordered correlators, important
as they often represent what can be physically measured, are accessed via
analytic continuations of temperature correlators.  In the next section
we consider multiple defects and in particular periodic arrays of defects.
In the final two sections, we consider a continuum
array of defects, first alone, then together with embedded impurities.

\newsec{Overview of Model}

\def\sp{S^+}
\def\sm{S^-}
\def\del{\partial}
\def\dm{\Delta}
\def\wo{\omega_0}
\def\wg{k}

In this section we describe the coupling between a single harmonic oscillator
defect
(placed at the origin) and a photon field in the environs of an optical fiber.
The oscillations of the defect have two possible physical sources.  The first,
going under the name ``atomic polarizability'', arises from the distortion
of charge within an atom.  One can crudely model this by representing the
atom as a charged shell tied to an immobile nucleus through a spring of
strength $K = m\wo^2$ (m is the mass of the shell and equal to $Zm_e$
where $Z$ is the atomic number and $m_e$ is the mass of an electron).
The shell then executes simple harmonic motion around the fixed 
nucleus.  Typically the energy scale of these oscillations is 
$\hbar\wo \sim 10$eV.

The second source comes with the name, ``displacement polarizability''.  Here
the defect is considered as a pair of oppositely charged ions that oscillate 
about their centre of mass.  Because these oscillations are phonon-like
(i.e. the heavier ions are oscillating and not the electronic shell),
their energy scale $\hbar\wo$ is typically $10^{-1}\sim 10^{-2}$eV.  In
this section we will consider the defect to be of this latter type.  At
energy scales of the former, we would also need to consider the 
atomic polarizability of the fiber medium in which the photons propagate and
in which the defects are embedded.  Instead we are able to treat the
medium of propagation as vacuum (or equally easily, a medium of static
dielectric constant, $\epsilon_0$).  We will delay the inclusion of the
dynamic properties of the medium to Section 7.

The oscillator, minimally coupled to the photon field in the radiation 
gauge,
has the Hamiltonian
\eqn\eIIii{
H =  {{\rm p}^2 \over 2m} + {m\wo^2 \over 2} \rm{q}^2 - 
{\rm{p} e \over 2m}{\bf{A}} (x=0) +{e^2 \over 2 m c^2} {\bf{A}}^2(x=0).}
By now introducing, in the standard way, the
variables
\eqn\eIIiii{\eqalign{
S^\pm &= \pm i\left( m\wo \over 2\hbar \right)^{1/2} {\rm q} + 
{1 \over \left( 2 m\hbar \wo \right)^{1/2}}{\rm p} ;\cr
1 &= \left[\sm ,\sp \right] ,}}
this Hamiltonian can be recast as
\eqn\eIIiv{
H = \hbar\wo \left(\sp\sm + 1/2\right) -
{e \over c} \left({\hbar \wo \over 2 m}\right)^{1/2} 
\left (\sp + \sm \right) {\bf A}(x=0) + {e^2 \over 2 m c^2}{\bf A}^2(x=0) .}
The interaction terms are then of two sorts: one coupling the photon field
to the oscillator and one giving the photon a ``mass'' at $x=0$.
Throughout our analysis we will keep this latter term.
The ${\bf A}^2$ term contributes equally to the quantities which we will be
interested in calculating.  Its omission, on the other hand, 
thereby violating gauge invariance, leads to unphysical results.
Fortunately, its inclusion does not
overly complicate the analysis.

As we ultimately want to derive equations of motion, we want to write the
above as an action.  Because $\sp$ and $\sm$ can be thought of as conjugate
variables, i.e. $\left[i\hbar\sm , \sp\right] = i\hbar$,
this can be easily done.  The action is
\eqn\eIIv{\eqalign{
S_{sho} &= \int dt \left( -i\hbar\sm\del_t\sp - H \right) \cr
&= \int dt \left( -i\hbar\sm\del_t\sp - \hbar\wo\sp\sm \right. \cr
& ~~~~~~~~~~~~~~~~~ \left. +
{e \over c}\left({\hbar \wo \over 2 m}\right)^{1/2}
\left( \sp + \sm \right) {\bf A}(x=0) - 
{e^2 \over 2 m c^2}{\bf A}^2(x=0)\right).}}
In writing this we have dropped the zero point energy of the oscillator.

We now focus our attention on the free dynamics of the photons.  The action
determining these dynamics is
\eqn\eIIvi{
S_A = -{1 \over 8\pi} \int d^4x F^{\mu\nu}F_{\mu\nu} .}
For a fiber, $S_A$ can be dimensionally reduced.  We begin by
setting the scalar potential $A_0$ to zero.  
We then take the fiber along the $\hat x-$direction.  
${\bf A}$ is necessarily in some transverse direction.
Assuming only one direction of ${\bf A}$ is excited, say ${\bf \hat n}$ where 
$\bf \hat n \cdot \hat x = 0$, we can write 
${\bf A} = A {\bf \hat n}$.  If $A$ is independent of the transverse 
directions, $\bf \hat y$ and $\bf \hat z$, $S_A$ reduces to
\eqn\eIIvii{
S_A = {D\over 8\pi}\int dt dx 
\left( c^{-2}(\del_t A)^2 - (\del_x A)^2\right),}
where D is the cross-sectional area of the fiber arising from performing
the integral $\int dy dz = D$.  We can rescale $A$ to turn $S_A$ into
an action describing a 2-d free massless boson:
\eqn\eIIviii{
A \rightarrow {\phi \over D^{1/2}}, }
so that 
\eqn\eIIvix{
S_A = {1\over 8\pi}\int dt dx 
\left( c^{-2}(\del_t \phi)^2 - (\del_x \phi)^2\right) . }
We list factors of $c$ and $\hbar$ so as to be able to easily provide
estimates of various couplings.  

The full action we consider is then
\eqn\eIIx{\eqalign{
S &= S_A + S_{sho} = {1\over 8\pi}\int dt dx 
\left( c^{-2}(\del_t \phi )^2 - (\del_x \phi )^2 \right) 
+ \int dt \left.\bigg( -i\hbar\sm\del_t\sp \right. \cr 
& ~~~~~~~~~~~~~~~~~~~~~~~~~~~~~ \left. -\hbar\wo\sp\sm 
+ (\hbar c)^{1/2} \lambda \left( \sp + \sm \right) \phi - 
{c \over \hbar} \dm^2 \phi^2\right).}}
where $\lambda$ and $\dm^2$ are
\eqn\eIIxi{\eqalign{
\lambda &= \left( \wo \over 2 m c D \right)^{1/2} {e \over c} ;\cr
\dm^2 &= {e^2 \hbar \over 2 m c^3 D}.}}
$\dm^2$ has units of mass and $\lambda$ has units of 
inverse length.  Because the bound, approximately by $10^{-4}$,
on the fiber diameter, $D$, is
large compared to atomic length scales, both $\lambda$ and
$\dm^2$ are small compared to the scale set by 
$\hbar \wo \sim 10^{-2} \rm{ev}$,
the typical oscillator energy.  We have
\eqn\eIIxii{\eqalign{
{\lambda \hbar c \over \hbar \wo} &\sim 10^{-7}, \cr
{\dm^2 c^2 \over \hbar \wo} &\sim 10^{-14}. }}
We thus see that the mass perturbation $\dm^2$ is on the same order as
$\lambda^2$, as required by gauge invariance.  
When  $\hbar = c = 1$ the two scales
are governed by 
\eqn\eIIxiia{
\lambda^2 = \Delta^2 \omega_0 .}
Thus $\Delta^2$ cannot be ignored in favour of $\lambda$, the
photon-oscillator coupling.  
Because of (2.12),
we will express all derived quantities such as scattering matrices,
band structure, correlators, etc., in terms of $\Delta^2$.

Though both couplings are small compared to
$\hbar\wo$, the perturbations are still able to lead to interesting physics.
It is possible to introduce other length scales into the problem by
considering multiple defects, which in combination with the existing
scales can lead to observable effects.  We will consider such cases in
Sections 6, 7, and 8.  But for now we continue to develop the single defect
formalism.

\newsec{Calculation of Defect S-Matrices}

Generally given a theory with particles $A(k)$, an integrable defect
perturbation produces scattering via the defect S-matrices, $T(k)$
and $R(k)$, as follows:
\eqn\eIIIi{
{\bf D}A_+(\wg ) = T(\wg )A_-(\wg ) {\bf D} + 
R(\wg ){\bf D}A_+(-\wg ) ,}
where {\bf D} is the defect operator.  {\bf D} is analogous to the
boundary operator {\bf B} discussed in \zam .
$A_\pm$ represents particles who live on the right ($x>0$)/left ($x<0$) side
of the defect.  
$T(\wg )$ is then just the amplitude that the particle, $A_+(\wg )$,
makes it across the defect, and
$R(\wg )$, the amplitude that it reflects back.

To derive $T$ and $R$ we reconsider the action in (2.9).  
Setting $\hbar = c = 1$ and 
transforming to Euclidean time, $t \rightarrow -it$,
this becomes
\eqn\eIIIiii{\eqalign{
S =& {1 \over 8\pi} \int dt dx \left( (\del_t\phi)^2 + (\del_x\phi)^2 \right) \cr
& ~~~~~~~~+ \int dt ~\wo \sp\sm - S^-\del_tS^+ - \lambda (\sm + \sp )\phi + 
\Delta^2\phi^2 .}}
By varying $\sm$, $\sp$, and $\phi$, we obtain the following equations 
of motion:
\eqn\eIIIiv{\eqalign{
0 &= \lambda \delta (x) (\sm + \sp ) + {1 \over 4 \pi}\left(\del_t^2\phi +
\del_x^2\phi\right) - 2 \delta(x)\dm^2\phi ;\cr
0 &= \delta (x)\left( -\lambda \phi + \wo S^\pm \mp \del_t S^\pm\right) . }}
The second of these two has the steady-state solution
\eqn\eIIIv{
S^\pm = \lambda \int^\infty_{-\infty} d\wg e^{kt} 
{\tilde\phi (\wg) \over \wo \mp \wg },}
where $\tilde\phi (\wg )$ is defined through
\eqn\eIIIvi{
\phi (x=0) = \int^\infty_{-\infty} d\wg e^{\wg t} \tilde\phi (\wg ).}
We can thus eliminate $S^\pm$ from the first equation of motion to obtain
\eqn\eIIIvii{
0 = 2\wo \lambda^2 \delta (x) \int^\infty_{-\infty} d\wg e^{\wg t}
{\tilde\phi (\wg) \over \wo^2 - \wg^2} + {1 \over 4\pi}\left( \del_t^2\phi
+\del_x^2\phi \right) - 2\dm^2\phi\delta (x) .}
To solve this equation we write $\phi$ as
\eqn\eIIIviii{
\phi (x) = \phi_+ (x) \theta (x) + \phi_-(x) \theta (-x) ,}
where $\phi_\pm$ are free massless bosons with mode expansions
\eqn\eIIIvix{
\phi_\pm (x) = \int^\infty_{-\infty} d\wg {1 \over (|\wg |)^{1/2}}
\left( A_\pm (\wg ) e^{-t|\wg | +ix\wg} + A^\dagger_\pm (\wg ) e^{t|\wg | -ix\wg} 
\right).}
We take $\phi(x=0)$ to be defined as $\phi(0) = 1/2(\phi_+(0) + \phi_-(0))$.
So our equation of motion can be written as
\eqn\eIIIx{
0 = \wo \lambda^2 \int^\infty_{-\infty} d\wg e^{\wg t}
{\tilde\phi_+ + \tilde\phi_- \over \wo^2 - \wg^2} +
{1 \over 4\pi}\left( \del_x\phi_+ (0) - \del_x\phi_-(0) \right) -
\dm^2(\phi_+(0) + \phi_-(0)).}
Second order derivatives have vanished because 
$(\del_t^2 + \del_x^2)\phi_\pm = 0$.

To derive $T(\wg )$ and $R(\wg )$, we divide the above equation into its 
`+' and `-' parts, say $f(A_+(\wg)) + g(A_-(\wg)) = 0 $, and so interpret it as
vanishing when acting on the defect operator in the following manner:
\eqn\eIIIxi{
{\bf D} f(A_+(\wg)) + g(A_-(\wg)){\bf D} = 0 .}
Note carefully the orderings of ${\bf D}$ and the $A_+$'s and $A_-$'s.  
Substituting the mode expansions into this equation together with the 
continuity equation
\eqn\eIIIxii{
{\bf D}\phi_+(0) = \phi_-(0){\bf D},}
fixes $T(\wg )$ and $R(\wg )$ to be
\eqn\eIIIxiii{\eqalign{
T(\wg ) &= {i(\wg^2 - \wo^2) \over i(\wg^2 -\wo^2) - 4\pi\dm^2\wg}, \cr
R(\wg ) &= {4\pi\dm^2\wg \over 
i(\wg^2 -\wo^2) - 4\pi\dm^2\wg} .}}
As a check on the validity of this result, we see $T(\wg )$ and $R(\wg )$
satisfy the defect unitarity conditions:
\eqn\eIIIxiv{\eqalign{
T(\wg )R(-\wg ) + T(-\wg )R(\wg ) &= 0 ;\cr
R(\wg )R(-\wg ) + T(\wg )T(-\wg ) &= 1.}}
These conditions are easily derived through applying \eIIIi\ twice, i.e.
allowing the particles to scatter through the defect twice.

The analytic structure of $T(\wg )$ and $R(\wg )$ is unsurprising.  Both
have poles at 
\eqn\eIIIxv{
\wg_p = \pm \wo (1 - {4\pi^2\Delta^4 \over \wo^2})^{1/2} -
{i2\pi\Delta^2}}
The real part of $\wg_p$ describes the energy splitting between two 
oscillator levels.  This splitting has been shifted slightly from $\wo$
in a process analogous to the Lamb shift.  Minus the imaginary part of $\wg_p$
equals the inverse lifetime of an excited state.  We then have
\eqn\eIIIxvi{
\tau^{-1} = -{\rm Im}\wg_p = {2\pi \Delta^2}. }
This lifetime is what is similar to what is found
in perturbative treatments using the master equation (see for
example \ref\wm{D.F. Walls, G.J. Milburn, {\bf Quantum Optics},
Springer-Verlag, New York, 1994, ch. 6.1).}).
This is unsurprising.
If we were to subject the theory to a perturbative analysis we would find
that the photon self-energy (i.e. the vacuum polarization) 
consists of a single diagram of lowest order and so $\tau^{-1}$ should
have no contributions higher than $\Delta^2$.  A similar situation
is found in perturbative analyses of the simple defect theories in 
\muss .

The interpretation of this pole differs from the standard one in integrable
Euclidean QFT.  Normally the imaginary part of $\wg_p$ would be interpreted as 
related to the energy of an excited oscillator state and the real part as
the decay time of the excited state.  But here the roles have been reversed,
as if we still were in Minkowski space.  This standard interpretation arises
from assuming that the pole in the S-matrix arises from an effective 
propagator of the form $(p^2 + m^2)^{-1}$.  But here this is not the case.
In Minkowski space, the pole appears as $\sim (\wg^2 - \wo^2)^{-1}$.
Both $\wg$ and $\wo$ are energies.  In going to Euclidean space, we then
have $\wg \rightarrow -i\wg$ and $\wo \rightarrow -i\wo$.  Thus the pole
structure retains its form in Euclidean space.

\newsec{Thermodynamics} 

Free photons in a cavity at a temperature $T$ are governed by
the well-known Planck distribution:
\eqn\Ti{
\rho (k) dk = \frac{dk}{2\pi} \inv{e^{k/T} -1} .}
The mean energy per unit length is given by
\eqn\Tii{
U(T) = \int_{-\infty}^\infty dk |k| \rho (k) = \frac{\pi}{6} T^2 .}
The above expression for $U(T)$ is the one-dimensional analog of the
Stephan-Boltzman law.  $U(T)$ can also be expressed as 
\eqn\Tiii{
U(T) = \frac{T^2}{L} \d_T \log Z, }
where $L$ is the length of the cavity, and
\eqn\Tiv{
\log Z = - \frac{L}{2\pi} \int_{-\infty}^\infty dk \log \( 1- e^{-|k|/T} \),}
is the free energy of the non-interacting photons.
We will now compute the modifications of the above blackbody theory due
to the presence of the  impurity.  We place the
impurity at the origin at $x=0$, let the cavity have length $L$,
$-L/2 < x < L/2$, and further impose periodic boundary conditions
at the ends of the cavity. 

For a single impurity,
we can  fold the system so that it is formulated as a theory on the
half line with the impurity on the boundary at $x=0$. 
This kind of folding was used in the work \ref\rlud{P. Fendley, 
A. W. W. Ludwig, H. Saleur., Phys. Rev. Lett 74 (1995) 3005.} in
their study of edge states in the $\nu =1/3$ quantum Hall system. 
Let $\phi = \phi_L + \phi_R$, where $\phi_L (x+t)$ and $\phi_R(x-t)$ are
the left and right-moving components of the massless scalar field
in the bulk.  Define even and odd fields as follows:
\eqn\Tivb{\eqalign{
\phi^e_L (x,t) &= \( \phi_L (x,t) + \phi_R (-x,t) \)/\sqrt{2}; \cr
\phi^o_L (x,t) &= \( \phi_L (x,t) - \phi_R (-x,t) \)/\sqrt{2}; \cr
\phi^e_R (x,t) &= \phi^e_L (-x,t); ~~~~~
\phi^o_R (x,t) =  - \phi^o_L (-x,t) . 
\cr }}
Also define $\phi^{e,o} = \phi^{e,o}_L + \phi^{e,o}_R$.  
Then the action \eIIx\ can be rewritten as a theory on the half-line
$x<0$: 
\eqn\Tv{\eqalign{
S &= {1\over 8\pi}\int^\infty_{-\infty} dt \int^0_{-\infty} dx 
\left(  (\d\phi^e)^2 + (\d\phi^o)^2 \right) \cr
& ~~~~~~~~~~~~ + \int^\infty_{-\infty}
dt \left.\bigg( -i\sm\del_t\sp - \right.  
 \left. \wo\sp\sm 
+  \frac{\lambda}{\sqrt{2}} \left( \sp + \sm \right) \phi^e  - 
 \frac{\dm^2}{2} ( \phi^e )^2\right).}}
Note that $\phi^e$ and $\phi^o$ decouple, and only $\phi^e$ couples 
to the impurity.  

\def\omo{\omega_0}

The interaction of the even photons with the impurity at the boundary
are now summarized by a reflection matrix $\CR_e$.  The algebra 
\eIIIi\ now only has a reflection piece:
\eqn\efz{
{\bf  B} A(k) = \CR_e(k) {\bf B} A(-k) , }
where ${\bf B}$ is a boundary operator. 
A similar computation to the one in section 3 gives
\eqn\Tvi{
\CR_e (k) = \frac{ i(k^2 - \omega_0^2) + 4 \pi \Delta^2 k }
{i(k^2 - \omega_0^2) - 4 \pi \Delta^2 k }
.}
A similar reflection matrix describes the odd photons.  However as the odd
photons do not couple to the defect, it is trivial, equaling unity.

\def\rt{\tilde{\rho_e}}

Thermodynamic properties can all be expressed in terms of $\CR_e$.
The quantization condition on the momentum $k$ of an even photon 
on the half-line of length $L/2$ is 
\eqn\Tvii{
e^{ikL} ~ \CR_e (k) = 1,       ~~~~~~k>0.}
We need only consider $k>0$ since it is implicit that the wave-function
satisfying the above quantization condition contains both a 
right-moving and reflected left-moving particle.  Taking the $k$-derivative
of the log of \Tvii\ yields
\eqn\Tviii{
\rt (k) = \inv{2\pi}  - \frac{i}{2\pi L} \d_k \log \CR_e(k), }
where $L\rt(k) dk$ is the number of allowed states between 
$k$ and $k + dk$.  Introducing a density of occupied levels
$\rho_e(k)$, one has for the partition function of the even field
\eqn\Tix{
\log Z_e = -\frac{U_e}{T} + \CS_e, }
where $U_e = L \int_0^\infty dk k \rho_e (k)$ and $\CS_e$ is the entropy:
\eqn\Tx{
\CS_e = L \int_0^\infty dk (\rho_e + \rt) \log (\rho_e + \rt) - \rho_e \log 
\rho_e - \rt \log \rt .}
Minimizing $\log Z_e$ with respect to $\rho_e$, one finds that the
Planck distribution \Ti\ for the even photons is modified to 
\eqn\Txi{
\rho_e(k) dk = \inv{e^{k/T} -1 } 
\( \inv{2\pi}  - \frac{i}{2\pi L} \d_k \log \CR_e(k) \) dk ,
}
and the even free energy becomes
\eqn\Txii{
\log Z_e = -L \int_0^\infty \rt (k) \log \( 1 - e^{-k/T} \). }
We now put this all together to find a total energy for the system.

The mean energy density separates into bulk and impurity contributions:
\eqn\Txiib{
U(T) = U_{\rm bulk} (T) + U_{\rm imp} (T).}
$U_{\rm bulk}$ is a sum of equal contributions from the even and odd sectors
of the theory,
\eqn\Txiiba{
U_{\rm bulk} = U^e_{\rm bulk} + U^o_{\rm bulk} = 2 \times \int^\infty_0
dk {k \over 2 \pi} {1 \over e^{k/T} - 1}.}
Hence $U_{\rm bulk} = \pi T^2 /6$.
$U_{\rm imp}$ is the piece from the even sector that scales as $L^{-1}$.
Hence
\eqn\Txiii{
U_{\rm imp} (T) = \frac{4 \Delta^2}{L}  
\int_0^\infty dk \frac{k}{e^{k/T} -1 } 
\frac{k^2 + \omo^2}{ (k^2 - \omo^2)^2 + (4\pi \Delta^2 k)^2 } .}
In the limit $\Delta \to 0$, the photons are decoupled from the 
impurity, and one expects $L U_{\rm imp}$ to be just the energy
of a decoupled harmonic oscillator:
\eqn\Txiv{
\lim_{\Delta \to 0} ~ L U_{\rm imp} = T^2 \del_T 
\log \big( \sum_{n=0}^\infty e^{-n\omo/T} \big)
= \wo \inv{e^{\omo/T} - 1} .}
We have set the ground state energy of the harmonic oscillator to zero
since in the scattering theory description the vacuum is defined to
have zero energy.  One can confirm that \Txiii\  has the proper
zero coupling limit, using the identity:
\eqn\Txv{
\lim_{\Delta^2/\omo \to 0} 
\inv{(k^2 - \omo^2)^2 + (4\pi \Delta^2 k)^2 } = \inv{8 \Delta^2 \omo^2}
\delta(k-\omo).}
Then,
\eqn\Txvi{
\lim_{\Delta \to 0} ~ L U_{\rm imp} (T) = \omo\inv{e^{\omo/T} -1},}
which agrees with \Txiv. 

In common physical situations, $\omo^2 \gg (4\pi \Delta^2 )^2$.  Making
this approximation in \Txiii, one can extract the following
low and high temperature behaviours:
\eqn\Txvii{
\eqalign{
U_{\rm imp} (T) &\approx \frac{2\pi^2}{3} \frac{\Delta^2}{L} 
\frac{T^2}{\omo^2} , ~~~~~ T\ll \omo ; \cr
U_{\rm imp} (T) &\approx \frac{T}{L}, ~~~~~ T \gg \omo . 
\cr
}}

\newsec{Computation of Correlators}

In this section we will consider how to calculate several types of
correlators that are useful in quantum optics problems.  
We begin with time-ordered correlators.

\def\td{\tilde{\bf{D}}}
\def\tt{\tilde{T}(\wg )}
\def\rt{\tilde{R}(\wg )}
\subsec{Time-Ordered Correlators}

There are two possible pictures in which to view the defect.  The first,
as in section 3, treats the defect as a point in space.  Particles then
scatter to and from the defect.  Because the Hamiltonian of the theory is
altered by the defect on any constant time surface, the particle spectrum
is altered.  This is reflected in the linear relations between the particle
creation/destruction operators as in 3.1 and 3.2.  In the second picture the
defect is a point in time, an initial condition.  Away from the defect
the Hamiltonian of the theory remains unchanged and so the particle
spectrum in the theory is the same.  This is of particular advantage in
calculating correlators.  The defect, as an operator, is expressible in
terms of the unperturbed Hilbert space and the correlator is then calculated
by inserting the defect within the {\it free} fields forming the correlator
in a time-ordered fashion:
\eqn\eVi{
\langle T(O_1(x_1) \cdots O_n(x_n)) \rangle = 
{\langle T(O_1^{\rm{free}}(x_1) \cdots \td O^{\rm{free}}_n(x_n)) \rangle
\over \langle \td \rangle} .}
On the l.h.s. of the above equation, the fields are the full Heisenberg fields
whereas the time evolution of the fields on the r.h.s. is governed by
the free Hamiltonian.  $\td$ 
(to be distinguished from the ${\bf D}$ of section 3) then is 
no more than the S-matrix for the theory.  It has a 
particularly simple form 
\eqn\eVii{
\td = \exp \bigg[ \int^\infty_{-\infty} 
d\wg \tt A^\dagger_+(\wg ) A_-(\wg ) + \int^\infty_0 \rt (A^\dagger_+(\wg )
A^\dagger_+(-\wg) + A_-(\wg ) A_-(-\wg ) ) \bigg].}
The $\pm$ subscripts on the photon operators indicate on which side of
the defect the particle lives: $+$ for $t>0$ and $-$ for $t<0$.  The
first term in the exponential transmits particles through the defect,
destroying a $-$ particle and replacing it with its $+$ counterpart.  Likewise
the latter two terms represent reflections off the defect.  Given $\td$'s
form, it is completely represented in terms of one particle scattering
amplitudes.  This is a consequence of the theory's 
integrability.  The form of the terms in the exponential are constrained
by momentum conservation; the total momentum must sum to zero.  Because all
the terms in the exponential commute with one another, there are no ordering
problems in representing $\td$.  We have assumed here that 
$\langle 0|\td | 0\rangle = 1$, that is the defect entropy is zero.  With
finite temperature this will not the case.

\def\hx{\hat{x}}
\def\ht{{\hat{t}\hskip.025in}}

To distinguish between the two pictures, space-time coordinates in the
first picture (the physical picture
with the defect as a boundary condition) will be written
as $(x,t)$ and the space-time coordinates in the second picture (with the
defect as an initial condition) by 
$(\hx , \ht )$.  The two are then related by $(t , x) = (\hx , \ht )$.
To compute the elements of $\td$, $\tt$ and $\rt$, one first expresses
the action in the rotated picture,
obtained from 3.3 via a Euclidean rotation and a reflection:
$x \rightarrow \hat{t}$ and $t \rightarrow \hat{x}$.
As before we can then
derive an equation of motion with the $S^\pm$ eliminated:
\eqn\eViv{
0 = \wo \lambda^2 \int^\infty_{-\infty} d\wg e^{i\wg \hx }
{\tilde\phi_+ + \tilde\phi_- \over \wo^2 + \wg^2} +
{1 \over 4\pi}\left( \del_{\ht} \phi_+ (0) - \del_{\ht} \phi_-(0) \right) -
\dm^2(\phi_+(0) + \phi_-(0)),}
where $\phi_+$ and $\phi_-$ are defined via 
$\phi = \theta (\ht )\phi_+ + \theta (-\ht ) \phi_-$.  As in section 3, we have
a continuity equation,
\eqn\eVv{
\phi_+ (\ht =0) = \phi_- (\ht =0).}
To interpret the action of the above 
two equations on $\td$, we divide each into
a part involving $\phi_+$ and a part involving $\phi_-$.  So both have
the form $f_+(\phi_+) - f_-(\phi_-) = 0.$  In terms of $\td$ these become
\eqn\eVvi{
f_+(\phi_+ ) \td g(\phi_- )|0\rangle = \langle 0|h(\phi_+) \td f_-(\phi_-).}
$g(\phi_-)$ and $h(\phi_+)$ represent arbitrary insertions of the fields
$\phi_-$ and $\phi_+$ between $\td$ and the vacuum state.
\eVvi\ tells us that as 
the fields $f_+(\phi_+)$ cross the defect $\td$, they are transformed
into $f_-(\phi_-)$.

To determine the form of $\tt$ and $\rt$, we expand out $\td$ to first order
in $\tt$ and $\rt$ and substitute into the above two equations.  Consistency
at the one particle level then yields
\eqn\eVvii{\eqalign{
\tt &= {\wo^2 +\wg^2 \over \wo^2 + \wg^2
+ 4\pi\Delta^2|\wg |},\cr
\rt &= {-4\pi\Delta^2 |\wg | 
\over \wo^2 + \wg^2 + 4\pi\Delta^2|\wg |}.}}
$\tt$ and $\rt$ can be related to the scattering matrices of section 3 via
an analytic continuation.  We see
\eqn\eVviii{
\tt (|\wg| \rightarrow -i\wg,\wg \rightarrow i|\wg| ) =  T(\wg ),}
and similarly with $\rt$.  This analytic continuation is effectively
a rotation of the defect back to the t-axis.  Here energies and momenta
are continued
via $E \rightarrow -ip$ and  $p \rightarrow iE$.

Knowing $\td$ we can now go on to calculate correlators.  We will compute
the two-point function of the photon field.
We have for this correlator,
\eqn\eVix{\eqalign{
\langle T (\phi (\hx ,\ht ) \phi (\hx ',\ht '))\rangle 
& = \theta (\ht - \ht ') \bigg[ \theta (\ht ) \theta (\ht ') 
\langle \phi (\hx ,\ht ) \phi (\hx ',\ht ')\td \rangle \cr
&~~~~~~~~~~~~~~~~ + \theta (\ht ) \theta (-\ht ') 
\langle \phi (\hx ,\ht ) \td \phi (\hx ',\ht ')\rangle \cr
&~~~~~~~~~~~~~~~~ + \theta (\ht ) \theta (-\ht ') 
\langle \td \phi (\hx ,\ht ) \phi (\hx ',\ht ')\rangle \bigg] \cr
& ~~~ + (\ht \leftrightarrow \ht ', \hx \leftrightarrow \hx ')}}
Each of the pieces of $\langle T (\phi \phi ) \rangle$ 
is easily calculated as $\phi\phi$ only couples
to the two particle contribution to $\td$.  So
\eqn\eVx{\eqalign{
\langle \phi (\hx ,\ht ) \phi (\hx ',\ht ')\td \rangle & = 
\langle \phi (\hx ,\ht ) \phi (\hx ',\ht ') \rangle_{\rm free} + 
2\int^\infty_0 d\wg {\rt \over \wg} e^{-(\ht +\ht ')|\wg |} 
\cos (\wg (\hx -\hx '));\cr
\langle \phi (\hx ,\ht ) \td \phi (\hx ',\ht ') \rangle & = 
\int^\infty_{-\infty} d\wg {\tt \over |\wg|} e^{-(\ht -\ht ')|\wg | 
+ i\wg (\hx -\hx ')};\cr
\langle \td \phi (\hx ,\ht ) \phi (\hx ',\ht ') \rangle & = 
\langle \phi (\hx ,\ht ) \phi (\hx ',\ht ') \rangle_{\rm free} + 
2\int^\infty_0 d\wg {\rt \over \wg} e^{(\ht +\ht ')|\wg |} 
\cos (\wg (\hx - \hx ')).}}
The correlator $\langle T (\phi \phi ) \rangle$
has now been calculated in the picture with the defect
as initial condition.  To obtain this same correlator in the physical
picture,
we rotate and reflect the system back via 
$\hx \rightarrow t$, $\ht \rightarrow x$.  If, for example, we supposed
$x = x' > 0$ in the original system, the correlator becomes
\eqn\eVxi{\eqalign{
\langle T (\phi (x,t) \phi (x , t')\rangle 
&= \langle \phi (\hx ,\ht ) \phi (\hx ',\ht )\td \rangle \cr
& = \langle \phi (t,x) \phi (t',x) \rangle_{\rm free} + 
2\int^\infty_0 {\rt \over \wg} e^{-2x|\wg |} \cos (\wg (t-t')).}}
The correlator 
only depends on $\rt$ as we have assumed 
that both space points are on the same side
of the defect.

\def\wg{\omega}
\subsec{Non-Time Ordered Correlators}

So far we have only considered time-ordered correlators.  But correlators
not so ordered can also be easily computed, a good thing as the correlators
in which one is often
interested for quantum optic calculations are not time-ordered.
For example, it common to want to compute
Glauber correlators, correlators of normal ordered products
of the creation and destruction pieces of the photon fields.  
In this section we will consider three examples of non-time ordered 
correlators: the photon density in the fiber at finite temperature, 
the spontaneous emission spectrum for an excited oscillator state, and the
correlators of the defect degrees of freedom.  We will do these calculations
in a finite temperature formalism.  It turns out the calculations are
easier to do in such a framework.  Euclidean correlators which are not
time ordered are not convergent for all of spacetime.  This makes the
rotation of the defect between viewing it as a boundary condition and
viewing it as an initial condition ill-defined for such correlators.
But in a finite temperature formalism, unordered correlators can be 
related to time-ordered temperature correlators for which the rotation,
$(x,t) \leftrightarrow (\hx ,\ht )$, poses no difficulties.

\def\hz{{\hat{z}}}
\def\hbz{{\hat{\bar{z}}}}
Temperature correlators are computed in much the same manner as the 
time-ordered correlators of the previous section.  The primary difference lies
in the mode expansions.  As we are at finite temperature, T, 
we expand the fields
on a cylinder of
circumference $\beta = 1/T =  4\pi l$:
\eqn\eVxxv{\eqalign{
\phi_l &= \sum_{n \geq 1} {1 \over n^{1/2}} ( \phi_n e^{-n\hz /l} 
+ \phi_{-n} e^{n\hz /l}) , ~~ \hz = (\ht +i \hx )/2 ;\cr
\phi_r &= \sum_{n \geq 1} {1 \over n^{1/2}} ( \bar\phi_n e^{-n{\hbz}/l} 
+ \bar\phi_{-n} e^{n{\hbz}/l}) , ~~ \hbz = (\ht -i \hx )/2 ;\cr
\phi &= \phi_l + \phi_r .}}
We have divided $\phi$ into its left and right moving components.  
Here the mode expansions 
are periodic in $\hx$ (and not $\ht$) as we intend to eventually
interchange time and space (so as to return to the original system).

\def\ttn{\tilde{T}(\wg_n )}
\def\rtn{\tilde{R}(\wg_n )}

The defect operator $\td$ for $T\neq 0$ is constructed in exactly the same 
fashion as in the $T = 0$ case.  We substitute the mode expansions
into the equations of motion and find
\eqn\eVxxvi{\eqalign{
0 &= {1 \over 4\pi}(\del_t \phi_+ - \del_t\phi_-) - \Delta^2 (\phi_+ + \phi_-)
+ \lambda (S^+ + S^-) ;\cr
S^\pm (\hx ) &= \lambda \sum_{n \geq 1} {1 \over n^{1/2}} \left[
{e^{i\wg_n \hx } \over \wo \mp i\wg_n} (\phi_{-n} + \bar\phi_n)
+ {e^{-i\wg_n \hx } \over \wo \pm i\wg_n} (\phi_n + \bar\phi_{-n})\right].}}
where the Matsubara frequencies are given by $\wg_n = n/2l$.  
From this we construct $\td$ as before:
\eqn\eVxxvii{
\td = g \exp\left[ \sum_{n \geq 1} \ttn (\phi^+_{-n}\phi^-_n + 
\bar\phi^+_{-n}\bar\phi^-_{-n} ) + \rtn (\phi^+_{-n}\bar\phi ^+_{-n}
+  \phi^-_n\bar\phi ^-_n) \right],}
where $\tilde{T}$ and $\tilde{R}$ 
are as in (5.6).
In the $T = 0$ case we were able to set $g$, the boundary
entropy to 1.  Here we cannot do so.  The partition function of the system is
\eqn\eVxxviii{
Z = \langle 0 | e^{\ht _1 H} \td e^{\ht _2 H} | 0 \rangle, }
where time extends to some $\ht _1$ and $\ht _2$ on either side
of the defect.  ($\ht _1 + \ht _2$ 
thus defines the length of the cylinder which we
are on.)  As $\ht _1,\ht _2 \rightarrow 0$,
we expect Z to reduce to its defect contribution.  Thus
\eqn\eVxxix{
Z_{imp} = \lim_{\ht _1,\ht _2 
\rightarrow 0} Z = \langle 0 | \td | 0 \rangle = g.}
Our TBA calculation from the previous section gives us
\eqn\eVxxx{
\log g =  - 4\Delta^2 \int^\infty_0 dk {(\wo^2 + k^2) \log (1 - e^{-k /T})
\over (\wo^2 - k^2)^2 + 16\pi^2 \Delta^4 k^2}.}
We point out that knowledge of g is not necessary for the computation of 
correlators as it is a constant which is always normalized away.

We now go on to compute the number of photons, $n(\wg )$, in the fiber.  
$n(\wg )$
is given in terms of the two point photon correlator:
\eqn\eVxxxi{
n(\wg ) = {\wg \over 2\pi} \int dt e^{-i\wg t} \langle \phi^c (x,t)
\phi^d (x,0) \rangle ,}
where $\phi^{c/d}$ are the creation/destruction pieces of the field.  These 
pieces are defined for non-free field theories to be (see 
\ref\glau{R. J. Glauber, Phys. Rev. 130 (1963) 2529.})
\eqn\eVxxxii{
\phi^{c/d}(x,t) = \pm {1\over 2\pi i}
\int dt' {\phi(x,t') \over t' - t \mp i\epsilon}.}
The factor $\wg / 2\pi$ in (5.17)
serves to remove the relativistic normalization
present in the mode expansion of the photon fields.  
$\langle\phi^c \phi^d \rangle$ is not time ordered and as indicated previously
would be
difficult to compute directly.  However by the fluctuation-dissipation
theorem we can reexpress this correlator in terms of a retarded Green's
function:
\eqn\eVxxxiii{\eqalign{
\int dt e^{-i\wg t} \langle \phi^c(x,t) \phi^d(x,0) \rangle
&= {2 \over e^{\beta \wg}
-1} {\rm Im} G_R(-\wg ) \theta (\wg ) ;\cr
G_R (\wg ) &= -i \int^\infty_0 dt e^{i\wg t} \langle 
[\phi(x,t),\phi(x,0)]\rangle .}}
This sort of identity between correlators
is easily proven by an explicit evaluation of
the thermal trace together with an insertion of states between the operators
(see \ref\rick{G. Rickayzen, {\bf Green's Functions and Condensed Matter},
Academic Press, NY, 1980.}).
The advantage of writing $n(\wg )$ in this form is that $G_R(\wg )$ is 
available directly as the analytic continuation of a temperature Green's
function,
\eqn\eVxxxiv{
G_R(\wg ) = G_T(-i\wg +\epsilon) = -\int^\beta_0 e^{i\wg_n t}
\langle \phi (x,t) \phi (x,0)\rangle_T \bigg| _{\wg_n = -i\wg + \epsilon},}
and $G_T$ is exactly what we are set up to calculate.

To compute $G_T = -\langle \phi (x,t) \phi (x,0)\rangle_T $ we go into the
rotated system.  Assuming $x>0$ we have
\eqn\eVxxxv{\eqalign{
G_T &= -\langle \phi (x,t) \phi (x,0)\rangle_t
= -\langle \phi_+ (\hx ,\ht ) \phi_+ (\hx ,0) \td \rangle \cr
& = -{1 \over l} \sum_{n \geq 1} {1 \over \wg_n} \cos (\wg_n t)
[1 + \rtn e^{-2\wg_n x}] .}}
Following the prescription above, we then find for $n(\wg )$,
\eqn\eVxxxvi{\eqalign{
n(\wg ) = {2 \theta (\wg) \over e^{\beta\wg} - 1} \bigg[ 1 &+ \sin (2\wg x)
{4\pi\Delta^2\wg (\wg^2 - \wo^2) \over 
(\wg^2 - \wo^2)^2 + 16\pi^2\Delta^4\wg^2} \cr
&  - \cos (2\wg x) {16 \pi^2 \Delta^4\wg^2 \over 
(\wg^2 - \wo^2)^2 + 16\pi^2\Delta^4\wg^2} \bigg] .}}
We see that we obtain the Planck distribution (times two - one for right
moving photons and one for left moving photons) plus a correction term.
This distribution should not be compared with the TBA calculation of the
previous section.  
The distribution there is not that of pure photons but rather of
their even and odd 
combinations that have been averaged over the length of the
system.

We now move on to calculating the spontaneous emission spectrum, $P( \wg)$,
arising from the decay of the oscillator from its first excited state.
As the electric field is given in terms of $\phi$ by $E \propto \del_t \phi$,
we define the power spectrum to be
\eqn\eVxxxvii{\eqalign{
P(\wg ) &=  {1 \over 4\pi^2} \int dt dt' e^{i\wg (t-t')} 
\langle S^- (0) \del_t \phi^c(x,t) \del_t \phi^d(x,t')S^+(0)\rangle \cr
&= {\wg^2 \over 4\pi^2} \int dt dt' e^{i\wg (t-t')} 
\langle S^- (0) \phi (x,t)\rangle 
\langle \phi (x,t')S^+(0)\rangle .}}
In the second line the four-point correlator has first been
factored into two two-point 
correlators and $\phi^{c/d}$ replaced with $\phi$, both possible
as $S^\pm$ only creates or destroys single particles, 
and then the resulting expression has been
integrated
by parts.  These steps are only possible at $T=0$.  However we will let
$T \neq 0$ in order to compute the two-point functions and then
take $T \rightarrow 0$ at the end.  We can
make a further simplification by noting that
\eqn\eVxxxviii{
\int  dt e^{-i\wg t} \langle S^- (0) \phi (x,t)\rangle 
= \bigg[ \int dt e^{i\wg t} \langle \phi (x,t)S^+(0)\rangle\bigg]^*.}
We thus need only  to compute $\langle \phi (x,t) S^+ (0)\rangle$.

As before we compute $\langle \phi (x,t) S^+ (0)\rangle$ 
through a temperature
correlator.  In this case we have
\eqn\eVxxxvix{\eqalign{
\int {dt \over 2\pi}  e^{i\wg t} \langle \phi (x,t) S^+ (0)\rangle 
&= {i \over 2 \pi} {\theta (\wg ) \over 1 - e^{-\beta \wg}}
(G_T(-i\wg + \epsilon ) - G_T(-i\wg - \epsilon )); \cr
G_T (\wg_n ) &= -\int^\beta_0 dt e^{i\wg_n t} \langle \phi (x,t) 
S^+ (0)\rangle_T.}}
Our expression involves both the retarded $G_T(-\wg + \epsilon )$
and the advanced $G_T(-\wg - \epsilon )$ Green's functions as the
combination $\phi S^+$ is not Hermitean.  Again this identity can be verified
by evaluating the thermal trace and inserting a complete set of states
between $\phi$ and $S^+$.  The temperature correlator is computed as
before by rotating the system.  Using (5.12) we then find,
\eqn\eVxl{\eqalign{
\langle \phi (x,t) S^+ (0)\rangle_T &= \langle \phi_+(\hx ,\ht )
S^+ (0) \td \rangle \cr
&= {\lambda \over 2l} \sum_{n \geq 1} {\ttn \over \wg_n}
\bigg[ {e^{-\wg_n(x+it)} \over \wo - i\wg_n} - 
{e^{-\wg_n(x-it)} \over \wo + i\wg_n} \bigg].}}
The Matsubara decomposition of this correlator is
\eqn\eVxli{
G_T (\wg_n )  = -{2\pi\lambda \over \wg_n}{1 \over \wo - i\wg_n}
(\theta (\wg_n) e^{-\wg_n x} \ttn - \theta(-\wg_n) e^{\wg_n x}
\tilde{T}(-\wg_n)).}
Putting everying thing together and taking $T \rightarrow 0$, 
we find the power spectrum to be
\eqn\eVxlii{\eqalign{
P(\wg )  & = {2\Delta^2 \wo \theta (\wg )\over (\wo - \wg )^2} 
{(\wo^2 -\wg^2)^2 \over D}\left[ 1
+ {\cos (2\wg x) \over D} ((\wo^2-\wg^2)^2 - 16\pi^2\Delta^4\wg^2) \right. \cr
& \left. ~~~~~~~~~~~~~~~~~~~~~~~~~~~~~~~~~~~~ - {\sin (2\wg x) 
\over D} (8\pi\Delta^2\wg (\wo^2-\wg^2))\right],\cr
D & = (\wo^2 -\wg^2)^2 + 16\pi^2\Delta^4\wg^2.}}
Typically the power spectra is computed for near resonance ($\wg \simeq \wo$).
Near resonance $P(\wg )$ reduces to
\eqn\eVxliii{
P(\wg) = \theta(\wg) {4\Delta^2\wo\sin^2(\wg x) \over (\wo-\wg)^2 + 
4\pi^2\Delta^4}.}
We see that $P(\wg )$ takes on the familiar Lorentzian lineshape with 
halfwidth equal to $4\pi\Delta^2$.  Near resonance the lowest
order term of $P(\wg)$ has the form predicted by classical theory 
(see \ref\aeb{L. Allen and J. H. Eberly, {\bf Optical Resonance and Two-Level
Atoms}, Dover, New York, 1987.})
and the same form predicted by first order perturbation theory employing
the master equation (see \wm ).  The spatial dependence in the power
spectra is an artifact of not including the power from the magnetic
field.  Doing so leaves $P(\wg )$ independent of $x$.

This basic Lorentzian lineshape is also shared by a two-level system
interacting with a photon field near resonance (see \wm ).  Where the
physics of the the two-level system disagrees with that of an SHO defect
is in the presence of an external driving field.  For a SHO defect the
response spectrum resulting from an external field is a $\delta$-function.
This is only true for a two-level system for a weak driving field.  As the
driving field intensity is increased, the $\delta$-function spectrum
of the two-level system broadens and then splits into three Lorentzians,
one centered at the driving frequency and two sidebands equidistant from
the central one.  The two systems diverge in behaviour in large external
fields as the two-level system saturates in its upper level for such field
whereas the SHO can keep being excited into ever higher levels.  At low 
driving intensities the SHO spends its time mostly its in ground state or
its first excited level and so mimics a two-level system.

As the final illustration of computing non-time ordered correlators, we 
compute the correlators of the defect degrees of freedom, i.e. 
$\langle S^{\gamma } (t) S^{\delta } (0) \rangle$, 
$\gamma , \delta = \pm$.  Even though the
defect degrees of freedom
have been integrated out, we can still compute their correlators as
we know how $S^\pm$ are related to the photon field.  As before 
$\langle S^{\gamma} (t) S^{\delta } (0) \rangle$ is related to 
a temperature correlator via
\eqn\eVxliv{\eqalign{
\langle S^\gamma S^\delta \rangle (\wg ) =
\int {dt \over 2\pi}  e^{i\wg t} \langle S^\gamma (t) S^\delta (0)\rangle 
&= {i \over 2 \pi} {\theta (\wg ) \over 1 - e^{-\beta \wg}}
(G_T(-i\wg + \epsilon ) - G_T(-i\wg - \epsilon )); \cr
G_T (\wg_n ) &= -\int^\beta_0 dt e^{i\wg_n t} \langle S^\gamma (t) 
S^\delta (0)\rangle_T.}}
$\langle S^{\gamma} (t) S^{\delta} (0) \rangle$
is computed by rotating the system
\eqn\eVxlv{
\langle S^{\gamma} (t) S^{\delta} (0) \rangle_T = \langle S^\gamma_+ (\hx )
\td S^\delta_- (0) \rangle_T .}
As $S^\gamma$ and $S^\delta$ are defined at $\hx = 0$, we have a choice as to
their ordering with respect to $\td$.  The expressions for the correlators
are independent of this choice (as they should be).  We have chosen
in the above
to place $S^\gamma$ and $S^\delta$ on either side of $\td$.  Using (5.12),
the correlator on the r.h.s. of the above is easily calculated:
\eqn\eVxlvi{
\langle S_+^{\gamma} (\hx ) \td S_-^{\delta} (0) \rangle_T =
\Delta^2 \wo \sum_{n \geq 1} {1 \over n} \bigg[ 
{e^{i\wg_n t} \over (\wo - i\gamma\wg_n)(\wo +i\delta\wg_n)} +
{e^{-i\wg_n t} \over (\wo + i\gamma\wg_n)(\wo - i\delta\wg_n)} \bigg].}
Then by \eVxliv\ , $\langle S^\gamma S^\delta \rangle (\wg )$ equals
\eqn\eVxlvii{
\langle S^\gamma S^\delta \rangle (\wg ) = 
{2 \Delta^2 \wo \theta (\wg ) \over \wg (1-e^{-\beta \wg})}
{(\wo^2 - \wg^2)^2 \over (\wo + \gamma\wg )(\wo - \delta\wg )}
{1 \over (\wo^2-\wg^2)^2 + 16\pi^2\Delta^4\wg^2}.}
One should notice that the correlators $\langle S^+ S^- \rangle$,
$\langle S^+ S^+ \rangle$, and $\langle S^- S^- \rangle$ do not
vanish.  This is a result of the coupling between the photon field and
oscillator.  Because the interaction is $(S^+ + S^-)\phi$ (and not
$S^+\phi^{d} + S^-\phi^{c}$), 
$S^+$ and $S^-$ have both creation and destruction
parts and so undergo vacuum fluctuations.

\newsec{Multiple Defects}

\def\wg{k}

The analysis of a single defect in the previous sections can be easily
extended to multiple defects.  We will first consider a set of $n$ defects
placed arbitrarily and then move on to a periodic array of defects.

\subsec{n Defects}

The action for a set of n defects is 
\eqn\eVIi{\eqalign{
S =& \int dt dx \bigg[ {1 \over 8\pi} \left( (\del_t\phi)^2 + 
(\del_x\phi)^2 \right) \cr
& ~~~~~~~~\left. + \sum^n_{i=1}
\delta (x-x_i) \left( -S_i^-\del_tS_i^+ + \omega_i \sp_i\sm_i - 
\lambda_i (\sm_i + \sp_i )\phi + \Delta_i^2\phi^2\right)\right],}}
where the i-th defect has been placed at $x_i$, $x_{i+1} > x_i > x_{i-1}$.  
We have allowed each defect
to have different coupling strengths to the photon field.  The equations of 
motion for this action are
\eqn\eVIii{\eqalign{
0 &= \sum^n_{i=1}\delta(x-x_i) \left[ \lambda_i (\sm_i + \sp_i ) 
- 2 \dm_i^2\phi\right]+ {1 \over 4 \pi}\left(\del_t^2\phi +
\del_x^2\phi\right);\cr
0 &= \delta (x-x_i)\left( -\lambda_i \phi + 
\omega_i S_i^\pm \mp \del_tS_i^\pm\right) . }}
By assuming $\phi$ has a free solution in the intervals in between the
defects,
\eqn\eVIiii{\eqalign{
\phi &= \phi_{n+1}\theta(x-x_n) + \phi_n\theta (x_n-x)\theta (x-x_{n-1})
+ \cdots \cr 
&~~~~~~~~~~~~~~~~~+ \phi_2\theta (x_2-x)\theta (x-x_1) + 
\phi_1\theta (x_1-x),\cr
0 & = (\del^2_t + \del^2_x)\phi_i, ~~~~~1 \leq i \leq n+1,}}
the above equations of motion can be reduced to 
\eqn\eVIiv{\eqalign{
0 = \sum^n_{i=1}\delta(x-x_i) \bigg[
\omega_i \lambda_i^2 \int^\infty_{-\infty} d\wg e^{\wg t} 
& {\tilde\phi_{i+1} (\wg) + \tilde\phi_i (\wg )
\over \wg^2 - \omega_i^2}  \cr
&  + \dm_i^2(\phi_{i+1} +\phi_i) - {1 \over 4\pi}\left( \del_x\phi_{i+1}
-\del_x\phi_i \right) \bigg] ,}}
where $\tilde\phi_i$ is, as before,
\eqn\eVIv{
\phi_i (x_i,t) = \int^\infty_{-\infty} dk e^{\wg t}
\tilde\phi (\wg ,x_i) .}
We have thus obtained a set of n uncoupled equations.  These can be solved
in same fashion as in Section 3.

Let ${\bf D}_i$ be the defect operator for the i-th defect.  We define the
transmission and reflection matrices across this defect by
\eqn\eVIvi{
{\bf D}_i A_{i+1}(\wg ) = T_i A_i (\wg ) {\bf D}_i + R_i(\wg ) 
{\bf D}_i A_{i+1}(-\wg ) ,}
where the $A_i (\wg )$'s arise from the mode expansion of $\phi_i$.
\eVIiv\ then fixes $T_i$ and $R_i$ to be
\eqn\eVIvii{\eqalign{
T_i (\wg ) & = T(\wg , \omega_i,\Delta_i) ,\cr
R_i (\wg ) & = e^{-2i\wg x_i} R(\wg , \omega_i,\Delta_i) ,}}
where $T(\wg , \omega_i,\Delta_i)$ and 
$R(\wg , \omega_i,\Delta_i)$
are the matrices in \eIIIxiii\ .  We see that $R_i$ has picked up a phase,
relative to $R$,
a result of a reflection away from $x=0$.

$T_i$ and $R_i$ describe the transmission and reflection across the defect
at $x_i$.  However the transmission and reflection amplitudes across all
$n$ defects is considerably more complicated; one has to consider multiple
reflections and transmissions as the photon traverses the set of defects.
To simplify the computation, we introduce a transmission matrix defined
via
\eqn\eVIviii{
\left( \eqalign{ A_{i+1} (\wg ) \cr A_{i+1} (-\wg )}\right) =
{\bf M}_i \left( \eqalign{ A_i (\wg ) \cr A_i (-\wg )}\right).}
Using \eVIvi\ and its conjugate,
\eqn\eVIix{
A_i(-\wg ){\bf D}_i = T_i (\wg ){\bf D}_i A_{i+1} (-\wg ) + R_i(\wg ) 
A_i(\wg ){\bf D}_i ,}
${\bf M}_i$ is found to be
\eqn\eVIx{
{\bf M}_i = \left[ \eqalign{ & T_i^{-1}(-\wg ) \cr & R_i(-\wg )T^{-1}_i(-\wg )}
~~~\eqalign{ & R_i(\wg )T^{-1}_i(\wg ) \cr & T_i^{-1}(\wg )} \right].}
We can then write down the transmission matrix across all the defects:
\eqn\eVIxi{
{\bf M_T} = {\bf M}_n \cdots {\bf M}_1}
The overall scattering amplitudes, $T_T$ and $R_T$ are defined by
\eqn\eVIxii{
{\bf D_T} A_{n+1}(\wg ) = T_T A_1 (\wg ) {\bf D_T} + R_T(\wg ) 
{\bf D_T} A_{n+1}(-\wg ) ,}
where ${\bf D_T} = {\bf D}_n \cdots {\bf D}_1$.
These amplitudes are then related to ${\bf M_T}$ via
\eqn\eVIxiii{
{\bf M_T} = \left[ \eqalign{ & T_T^{-1}(-\wg ) \cr & R_T(-\wg )T^{-1}_T(-\wg )}
~~~\eqalign{ & R_T(\wg )T^{-1}_T(\wg ) \cr & T_T^{-1}(\wg )} \right].}

As it stands ${\bf M_T}$ is not computable analytically (though certainly
numerically) for an arbitrary set of n $x_i$'s, $\Delta_i$'s, 
and
$\omega_i$'s.  As such we will only consider a particularly simple case:
the case of two defects.  Taking $x_1 = 0$,
$T_T (\wg )$ and $R_T(\wg )$ can the be computed
to be
\eqn\eVIxiv{\eqalign{
T_T(\wg ) & = {T_1(\wg ) T_2(\wg ) \over 1 - R_1 (\wg) R_2 (\wg )} =
T_1(\wg ) T_2(\wg ) \sum^\infty_{n=0} (R_1 (\wg ) R_2 (\wg ) )^n \cr
R_T(\wg ) & = R_2(\wg ) + {R_1(\wg )T^2_2(\wg ) \over
1 - R_1(\wg )R_2(\wg )} = R_2(\wg ) + R_1(\wg )T^2_2(\wg ) \sum^\infty_{n=0}
(R_1 (\wg ) R_2 (\wg ) )^n .}}
The latter equality in the above two equations shows that $T_T$ and $R_T$
can be both computed through adding up multiple reflections and transmissions
as if this problem was a two interface problem in classical optics.

As with $T(\wg )$ and $R(\wg )$ in Section 3, both $T_T$ and $R_T$ exhibit
poles as a result of the oscillator defects.  
$R_T$ has simple poles at 
\eqn\eVIxv{
\wg_p = \pm \omega_2 (1 - {4\pi^2 \Delta^4_2 \over \omega_2^2})^{1/2}
- i 2 \pi\Delta_2^2 , }
indicative of reflection off the second defect.  There is no similar pole
for the first defect as we are scattering from the right.  If we were 
scattering from the left, a pole in the variables $\omega_1$ and $\Delta_1$
would appear in $R_T$.  $R_T$ and $T_T$ have another set of poles dependent
upon $x_2$ representing a resonance arising from interactions between
the two defects.  In the limit $x_2 = 0$ and $(\Delta_2,\omega_2) = 
(\Delta_1,\omega_1)$, these poles appear at
\eqn\eVIxvi{
\wg_p = \pm \omega_1 (1 - {16\pi^2 \Delta^4_1 \over \omega_1^2})^{1/2}
- i 4 \pi\Delta_1^2 , }
In this case, the inverse lifetime has doubled, being the sum of the inverse
lifetimes of the two oscillators considered separately.  It sums 
for if either of the two oscillators emits a photon, the excited state
will decay.  

\subsec{Band Structure}

In this section we consider a periodic array of defects.  With such a
periodic structure, we expect the photon spectrum to be described by 
a band structure.  Rather than compute the band structure by
extracting it from ${\bf M_T}$, we
will be able to compute it through Bloch's theorem.

We begin with the Hamiltonian describing a three dimensional 
array of defects in
the fiber and their interaction with a photon field:
\eqn\eVIxv{
H = H_{\rm field} + \sum_{i,j,k}
\left( \wo \sp_{ijk}\sm_{ijk} +
\lambda (\sm_{ijk} + \sp_{ijk} )\phi (x_i,0,0) + \Delta^2\phi^2(x_i,0,0)
\right).}
Here $\{i,j,k\}$ marks out the locations of the defects in the array.
This array is three dimensional.  However the problem is still one dimensional
as $\phi$, as before, does not depend on the transverse coordinates, $y$ and
$z$.
To reduce this Hamiltonian to one
dimension, we define a collective oscillator representing a sum of all the
oscillators in a plane at $x_i$ transverse to the direction of the fiber,
\eqn\eVIxvi{
S^\pm_i = {1 \over \sqrt{N}} \sum_{jk} S^\pm_{ijk},}
where N is the number of oscillators in this plane.  
$S^\pm_i$ has the expected commutator
\eqn\eVIxvii{
[S^-_i,S^+_i] = 1.}
With this commutation relation, the above Hamiltonian can be recast as
\eqn\eVIxvii{
H = H_{\rm field} + \sum^{\infty}_{i=-\infty}
\left( \wo \sp_{i}\sm_{i} - 
\sqrt{N}\lambda (\sm_{i} + \sp_{i} )\phi (ia) + N\Delta^2\phi^2(ia)
\right).}
where $a$ is the lattice spacing of the array along the fiber.
(We do not need to assume the lattice spacing in the planes transverse to
the fiber is also $a$.)
We see both couplings have been rescaled, $\lambda$ by $\sqrt{N}$ and
$\Delta^2$ by $N$.  
The corresponding Minkowski action is 
\eqn\eVIxviii{\eqalign{
S =& \int dt dx \bigg[ {1 \over 8\pi} \left( (\del_t\phi)^2 -
(\del_x\phi)^2 \right) \cr
& \left. + \sum^\infty_{i=-\infty}
\delta (x-x_i) \left( -iS_i^-\del_tS_i^+ - \wo \sp_i\sm_i + 
\sqrt{N}\lambda (\sm_i + \sp_i )\phi - N\Delta^2\phi^2\right)\right].}}
For a lattice spacing 
in the transverse planes of $10^{-7}$cm and
$D \sim 10^{-4} {\rm cm}^2$, we have $\lambda /\wo \sim 10^{-2}$.  This is 
large enough to produce observable results.  

From \eVIxviii\ we can immediately write down the 
equations of motion governing such a periodic array of defects:
\eqn\eVIxv{\eqalign{
0 = \sum^\infty_{n=-\infty}&\bigg[ 
\wo N \lambda^2 \int^\infty_{-\infty} d\wg e^{i\wg t} 
{\tilde\phi_n (\wg,na) + \tilde\phi_{n-1} (\wg,(n-1)a)
\over \omega_0^2 - \wg^2}  \cr
&  - N \dm^2(\phi_n(na) +\phi_{n-1}((n-1)a)) +
{1 \over 4\pi}\left( \del_x\phi_n(na)
-\del_x\phi_{(n-1)}(na) \right) \bigg] ,}}
where $a$ is the spacing between the defects.
We have written $\phi$ as
\eqn\eVIxvi{
\phi (x) = \sum^\infty_{n=-\infty} \phi_n(x) \theta (na-x)\theta (x-(n-1)a).}
Each $\phi_n$ is free and so satisfies $(\del_t^2 - \del_x^2)\phi_n = 0$.

To solve this equation it is necessary
to uncouple the various $\phi_n$'s from one
another.  We do this via Bloch's theorem.   It tells us 
the solutions to these equations must satisfy
\eqn\eVIxvii{
\phi_n(na,t) = e^{ipa}\phi_{n-1}((n-1)a,t) .}
This constraint, together with the continuity condition,
\eqn\eVIxviii{
\phi_n (na,t) = \phi_{n-1}(na,t),}
are enough to specify the band structure.  After some algebra, we find
the consistency of (6.20), (6.22), and (6.23), demands
\eqn\eVIxix{
\cos (p a) = \cos (\wg a) + {4\pi N\Delta^2\wg
\over \wg^2 - \wo^2} \sin (\wg a).}
Here $p$ is to be understood as the photon momentum and $\omega = |\wg|$ 
as the photon
energy.  

\centerline{\psfig{figure=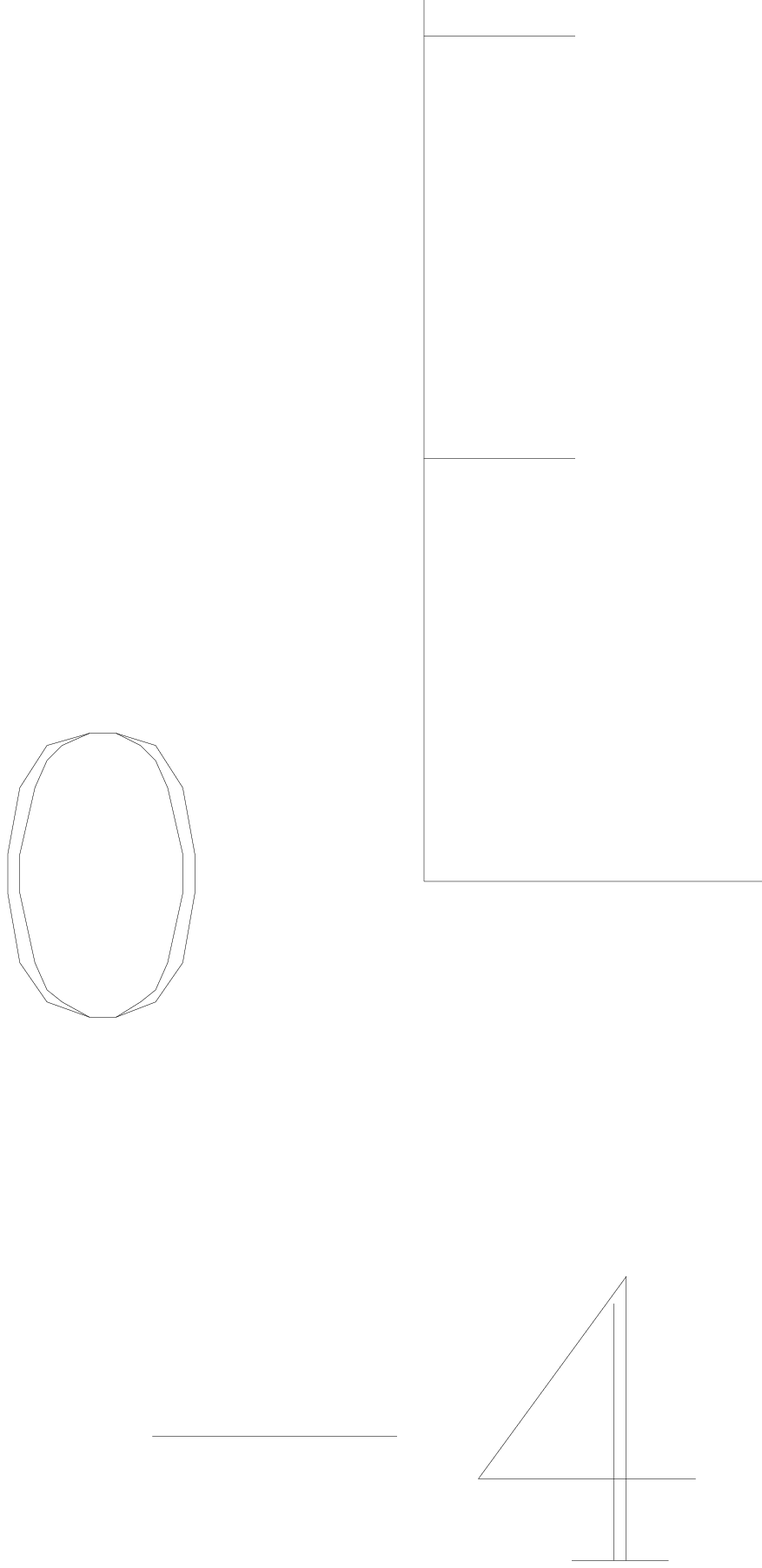,height=4in}}
\indent\vbox{\noindent\hsize 5in Figure 1: 
A plot of the band structure for a periodic array of defects.
We have taken 
$\wo^2 a^2 = 2$ and $4\pi N \wo^2\Delta^2 a^3 = 1$.
We have exagerated the value of $\Delta$ 
(or alternatively, underestimated $\wo a$) in order to make the
gaps visible.}

\vskip .20in
A plot illustrating the band structure is given in Figure 1.  
In this plot are drawn the first two bands and a portion
of the third in the reduced zone scheme.  There are three notable features
to the band structure: 1) in the lowest band, the energy, $\wg$, approaches
$0$ as $k \rightarrow 0$;
2) there is a gap between the first and the second gap, a result of
the resonance at $\wg = \wo$; and 3) there 
is a smaller gap between the second and
third bands, more typical as it appears as a small perturbation about
the edge of the first Brillouin zone.  

The existence of a mode at $\omega = k = 0$ results from gauge invariance.
Because of the simplifications introduced
by $\wo \Delta^2 = \lambda^2$, the numerator in (6.26)
vanishes as $\wg \rightarrow 0$.  If the ${\bf A}^2$ term had been omitted
from the action,
we would have found that long wavelength, low energy photons cannot propagate,
as if the photons had an imaginary mass.  One, however, 
can ignore the ${\bf A}^2$ term if, at the same time, one makes the dipole
approximation and couples the oscillators to the {\bf E}-field in place of
the vector potential, ${\bf A}$.  
This restores gauge invariance and so leads to the correct physics.

The effect of the pole at $\wg =\wo$ in \eVIxix\ is to divide what
would be the first unperturbed band $\wg = |k|$, $-\pi /a \leq k \leq \pi /a$
into two bands.
The gap between these two bands, if small, 
is given by
\eqn\eVIxxi{
\Delta E = \bigg| {4\pi N\Delta^2 \over \sin (\wo a)}\bigg| .}
The division into two branches is similar to what one sees in the dispersion
relation for photons in an covalent crystal (see \ref\am{N. Ashcroft and
D. Mermin, {\bf Solid State Physics}, W. B. Saunder Company, Philadelphia,
1976, ch. 27.} - we will derive this dispersion relation in the next
section).  However there are two notable differences in \am 's
treatment from ours.  There, the defects are coupled to the
E-field and not to the vector potential (the dipole approximation)
and the ${\bf A}^2$ term is ignored.  
The second difference is that the pole in \am~is shifted
away from $\wg = \wo$.  It seems plausible that this is a result of not
treating all the oscillators in unison.  
When we treat an isolated defect in Section 3, we see a similar shift in
the pole away from $\wo$ in $T(\wg )$ and $R(\wg )$.

The last feature of the band structure that bears comment is the
gap between the second and third bands.  This gap, far smaller than the
previous one, is equal to
\eqn\eVIxxii{
\Delta E = \bigg|{8\pi^2\Delta^2 N\over \pi^2 - \wo^2a^2} \bigg|.}
We have ignored higher order terms in $\Delta^2$.
Similar, but smaller, gaps will appear between the higher order bands.  Again 
to lowest order in $\Delta^2$, the
gap opening up at $\wo a = n\pi$ will equal (provided $\wo \neq n\pi/a$)
\eqn\eVIxxiii{
\Delta_n E = \bigg|
{8 N \pi^2\Delta^2 
n \over n^2\pi^2 - \wo^2a^2} \bigg|.}
Such gaps are what one typically expects to find in the case of small 
perturbations: the band structure is only altered around the edges of the
Brillouin zones.  However there is a difference from the standard case:
the opening up of the gaps is asymmetric about the energy $w = n\pi /a$:
the two solutions to \eVIxix\ defining the gap
are not $w = n\pi / a \pm \Delta_n E/2$
(symmetric splitting) 
but $w = n\pi /a$ and 
$w = n\pi/a + {\rm sgn}(n^2\pi^2 -\wo^2a^2)\Delta_n E$.

\newsec{Continuum of Defects}

\def\wg{\omega}

In this section we develop a theory describing a continuum set of defects
living along the entire length of the fiber.  As our starting point, we
take a set of point 
defects distributed uniformly with linear density $\rho$ along the fiber.  The
corresponding action is then
\eqn\eVIIi{\eqalign{
S =& \int dt dx \bigg[ {1\over 8\pi}\left( (\del_t\phi)^2 -
(\del_x\phi)^2 \right) \cr
& ~~~~~\left. + \sum_k
\delta (x-x_k) \left( -iS_k^-\del_tS_k^+ - \wo \sp_k\sm_k +
\lambda \sqrt{N}(\sm_k + \sp_k )\phi - N\Delta^2\phi^2(x_k)\right)\right].}}
We suppose the reduction to two dimensions, as in (6.17)-(6.20), has already
been made.
To take the continuum limit of this action, we introduce the field
variables $S^\pm (x)$:
\eqn\eVIIii{
S^\pm (x) = {1 \over \sqrt{\rho}} \sum_k S^\pm _k \delta (x-x_k).}
These then obey the expected commutation relations
\eqn\eVIIiii{
[ S^-(x) , S^+ (x')] = \delta (x-x') ,}
and the action may be rewritten as 
\eqn\eVIIiv{\eqalign{
S =& \int dt dx \bigg[ {1 \over 8\pi}\left( (\del_t\phi)^2 - 
(\del_x\phi)^2 \right) - iS^-(x)\del_tS^+(x) - \wo \sp (x)\sm (x)  \cr
& ~~~~~ +\sqrt{N\rho}
\lambda (\sm (x) + \sp (x) )\phi - N\rho\Delta^2\phi^2(x)\bigg].}}
In doing so, we have smoothed the defect distribution
by using the equality $\sum_k \delta (x-x_k) = \rho$,
a trick of Dirac's.
As $\rho$ is typically 
$\sim 10^{6}{\rm cm}^{-1}$, and $N \sim 10^{12}$, we have
$N\rho\Delta^2/\wo^2 \sim 10^{-4}$.
This, again, is
large enough to produce observable consequences.

From this action we are easily able to derive the exact dispersion relation.
As in previous sections, we can write down an equation of motion involving
$\phi$ alone:
\eqn\eVIIv{
0 = 2\wo^2\Delta^2 N\rho \int^\infty_{-\infty} d\wg e^{i\wg t} 
{\tilde\phi (\wg,x) \over \wg^2 - \omega_0^2} + 2N\rho\dm^2\phi (x) +
{1 \over 4\pi}\left( \del^2_t\phi -\del^2_x\phi \right) .}
The mode expansions of $\phi$ take the form
\eqn\eVIIvi{
\phi (x) = \int^\infty_{-\infty} dk {1 \over \sqrt{\wg (k)}}
\left( A(k) e^{-i\wg (k)t + ikx} + A^\dagger (k) e^{i\wg(k)t -ikx}\right)}
Substituting this into the equation of motion gives us the dispersion
relation
\eqn\eVIIvii{
\wg^2 (k) = 8\pi N\rho \dm^2 + k^2 + {8\pi N\rho\Delta^2 \wo^2 \over \wg^2(k) - \wo^2}.}
Like the band structure before, this has two branches.  They are described by
\eqn\eVIIviii{
\wg^2_\pm (k)  = {1 \over 2} \left[ (\wo^2 + 8\pi N\rho\dm^2 +k^2) \pm
\left( (k^2 + 8\pi N\rho 
\dm^2 - \wo^2)^2 + 32\pi N\rho \Delta^2\wo^2\right)^{1/2}\right].}
A plot of these two branches is found in the figure below:

\centerline{\psfig{figure=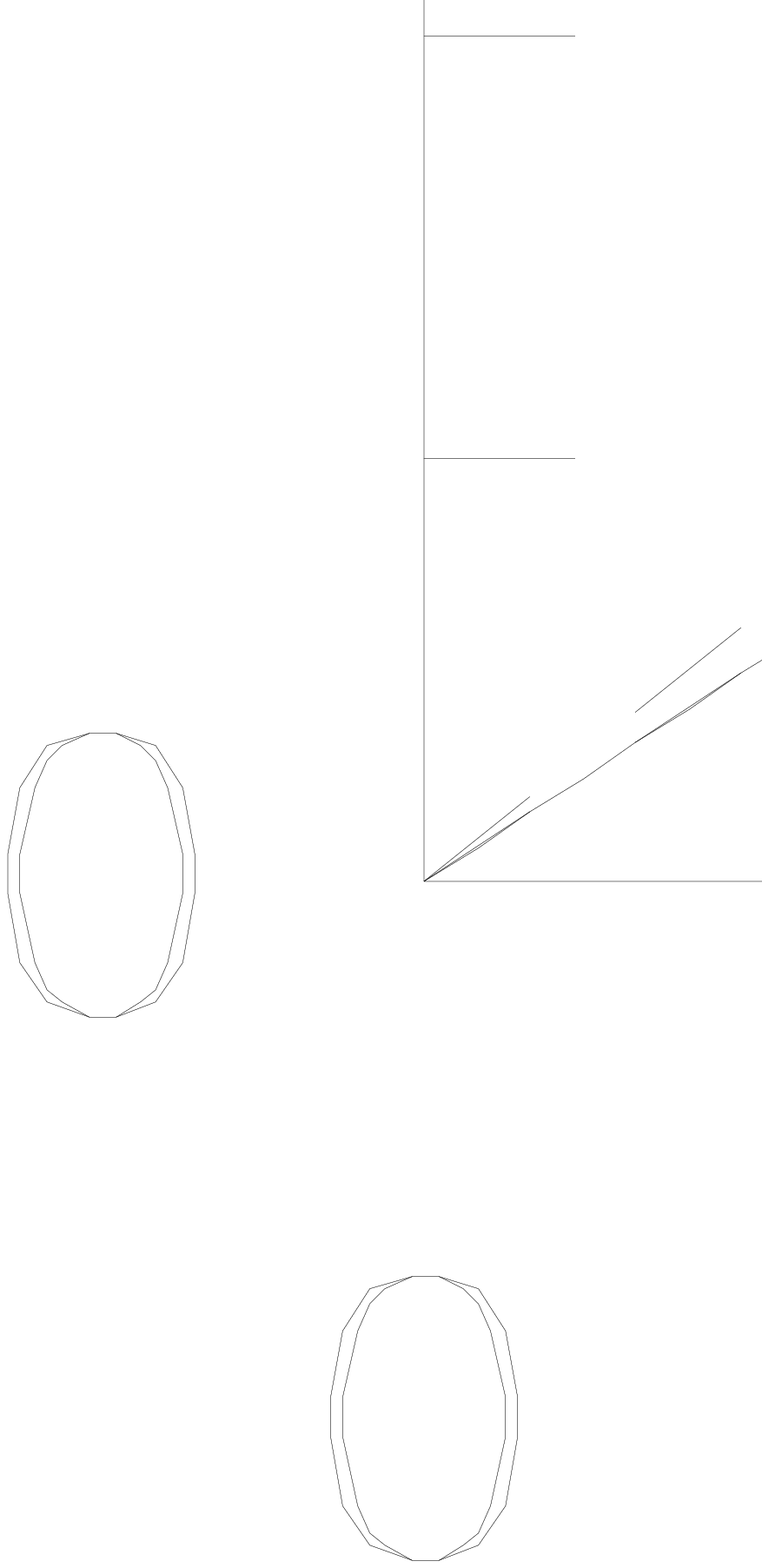,height=4in}}
\indent\vbox{\noindent\hsize 5in Figure 2: 
A plot of the dispersion relation for polaritons.  Both $w$ and $k$
are plotted 
in units of $\wo$.  
For this plot, we have taken 
$16\pi N \rho \Delta^2 \wo^{-2}= 1$.}

\vskip .2in
\noindent The modes described by these two branches of the 
dispersion relation are often referred to as polaritons.

As $k \rightarrow \infty$ both branches obtain asymptotic forms,
as indicated by the dashed lines.
The upper branch becomes photonic, $\wg = |k|$, as expected.  At high
energies the photons should pass undisturbed through a medium governed
by far lower energy scales.  The lower branch, on the other hand,
approaches a constant, $\wg = \wo$, in this limit, becoming like an
optical phonon.
As $k \rightarrow 0$, the upper branch approaches a constant,
\eqn\eVIIix{
\wg_+ (0)= \sqrt{\wo^2 + 8\pi N \rho \Delta^2}.}
A small gap thus separates these two branches. 
It is given by
$\Delta E = \sqrt{\wo^2 + 8\pi N\rho \Delta^2} - \wo$.
The lower branch, in contrast, becomes linear as $k \rightarrow 0$
and so describes photons
propagating in a medium of dielectric constant
\eqn\eVIIx{
\epsilon (0) = 1 + {8\pi N\rho \Delta^2 \over \wo^2}.}
Thus if we know $\epsilon (0)$, say experimentally, we can compute the 
effective coupling strength $N\rho\Delta^2$.

We are also able to compute the frequency dependent dielectric constant
that gives rise to these dispersion relations.   
From \eVIIviii\ and $\ep (\wg ) \wg^2 = k^2$ we have
\eqn\eVIIxa{
\epsilon (\wg ) = 1 + {8\pi N \rho \Delta^2 \over \wo^2 - \wg^2 }.}
The dielectric constant, 
$\epsilon (\wg )$ approaches $\pm \infty$ as
$\wg \rightarrow \wo$.  This asymptotic behaviour results in the two
separate branches of polaritons.  As $\wg \rightarrow \infty$,
$\epsilon (\wg )$ approaches $1$: as discussed previously, at energies
far greater than $\wo$ the photons cease to see the oscillators.

\newsec{Impurities in the Presence of a Continuous Medium}

In previous sections, 
we have treated the defects as if embedded in the vacuum.  The medium
in between the defects has been assumed to have no dielectric properties,
i.e. its dielectric constant, $\epsilon (\wg)$, has been set equal to $1$.  
This is not
necessarily a bad approximation.  The material out of which a fiber
is constructed typically has a dielectric constant with little frequency
dependence, i.e. $\epsilon (\wg ) = \epsilon_0$, for a wide range of
frequencies, $\hbar\wo \leq 10$eV.  So by a judicious insertion of 
$\epsilon_0$ in the equations from previous sections, 
we can take into account the medium
of propagation.  However if we wish we can do better.  As we have developed
here a theory of a continuous medium, it is possible
to solve the problem of a defect embedded in a medium with a frequency 
dependent dielectric constant.  We will consider the defect to arise
from atomic polarization.  Alternatively we could consider the defect
to be ionic in nature and suppose the natural frequency of the medium
to be far below the energy scale where the ionic
defect will also experience atomic
polarization.

\def\md{\Delta_d}
\def\mm{\Delta_m}
\def\wd{\omega_d}
\def\wm{\omega_m}
\def\ld{\lambda_d}
\def\lm{\lambda_m}

The action describing such a defect is a sum of medium's action and the 
defect's action:
\eqn\eVIIxi{\eqalign{
S & = S_{\rm med} + S_{\rm d} \cr
& = \int dt dx \bigg[ {1 \over 8\pi}\left( (\del_t\phi)^2 -
(\del_x\phi)^2\right) - iS^-(x)\del_tS^+(x) - \wm \sp (x)\sm (x)  \cr
& ~~~~~~~~~~ +\sqrt{N_m\rho_m}\lm (\sm (x) + \sp (x) )\phi -  
N_m\rho_m\mm^2\phi^2(x)\bigg] \cr
& ~~~~~~~~~~- \int dt \left[ iS^-\del_tS^+ + \wd \sp\sm  -\ld (\sm + \sp )\phi 
+ \md^2\phi^2\right].}}
We have labeled the defect and medium couplings with $d$ and $m$ subscripts
respectively.  The only difference in solving this theory from previous ones
is in the nature of the mode expansions.  Here the mode expansions are done
in terms of polaritons:
\eqn\eVIIxii{
\phi(x) = \sum_{j=+,-} \int^{\infty}_{-\infty} dk {1 \over \sqrt{w_j(k)}}
\left( A_j (k) e^{-i\wg_j(k)t + ikx} + A^\dagger_j(k)e^{i\wg_j(k)-ikx}\right).}
The sum reflects that the dispersion relation has two branches.  So the
scattering matrices for the polaritons off the defect are similar to those of
\eIIIxiii\ with $\wg (k) = |k|$ replaced by $\wg (k) = \wg_\pm (k)$ 
(where $\wg_\pm$ are given from \eVIIviii\ by replacing 
$\wo$ with $\wm$, $N\rho$ with $N_m\rho_m$, and $\dm$ with $\mm$).
Thus we have
\eqn\eVIIxiii{\eqalign{
T_\pm (k) &= {ik (\wg^2_\pm (k) - \wd^2) \over
ik(\wg^2_\pm (k) - \wd^2) - 4\pi\md^2\wg^2_\pm (k)};\cr
R_\pm (k) &= {4\pi\md^2\wg^2_\pm (k) \over
ik(\wg^2_\pm (k) - \wd^2) - 4\pi\md^2\wg^2_\pm (k)}.}}
We now need to index $T$ and $R$ to indicate which branch the polariton is
on.

Similarly we can imagine an array of defects imbedded in the continuous medium.
The band structure is then derived from \eVIxix\ by replacing the free
dispersion relation with $\wg_\pm$:
\eqn\eVIIxiv{
\cos (p_\pm a) = \cos (ka) + {4\pi N_d\md^2\wg_\pm^2(k)
\over k (\wg_\pm^2(k) - \wd^2)} \sin (k a),}
where $N_d$ is the number of defects in a section transverse to the fiber.
Here $p_\pm$ is the crystal momentum for polaritons on the upper/lower branch
of the dispersion relation.  For a given $p_\pm$, a set of $k$'s, $\{ k_i \}$,
satisfy this equation.  The energies corresponding to this $p_\pm$ are
then $\{ \wg_\pm (k_i)\}$.  A plot of the band structure is given in Figure 3
below.

\centerline{\psfig{figure=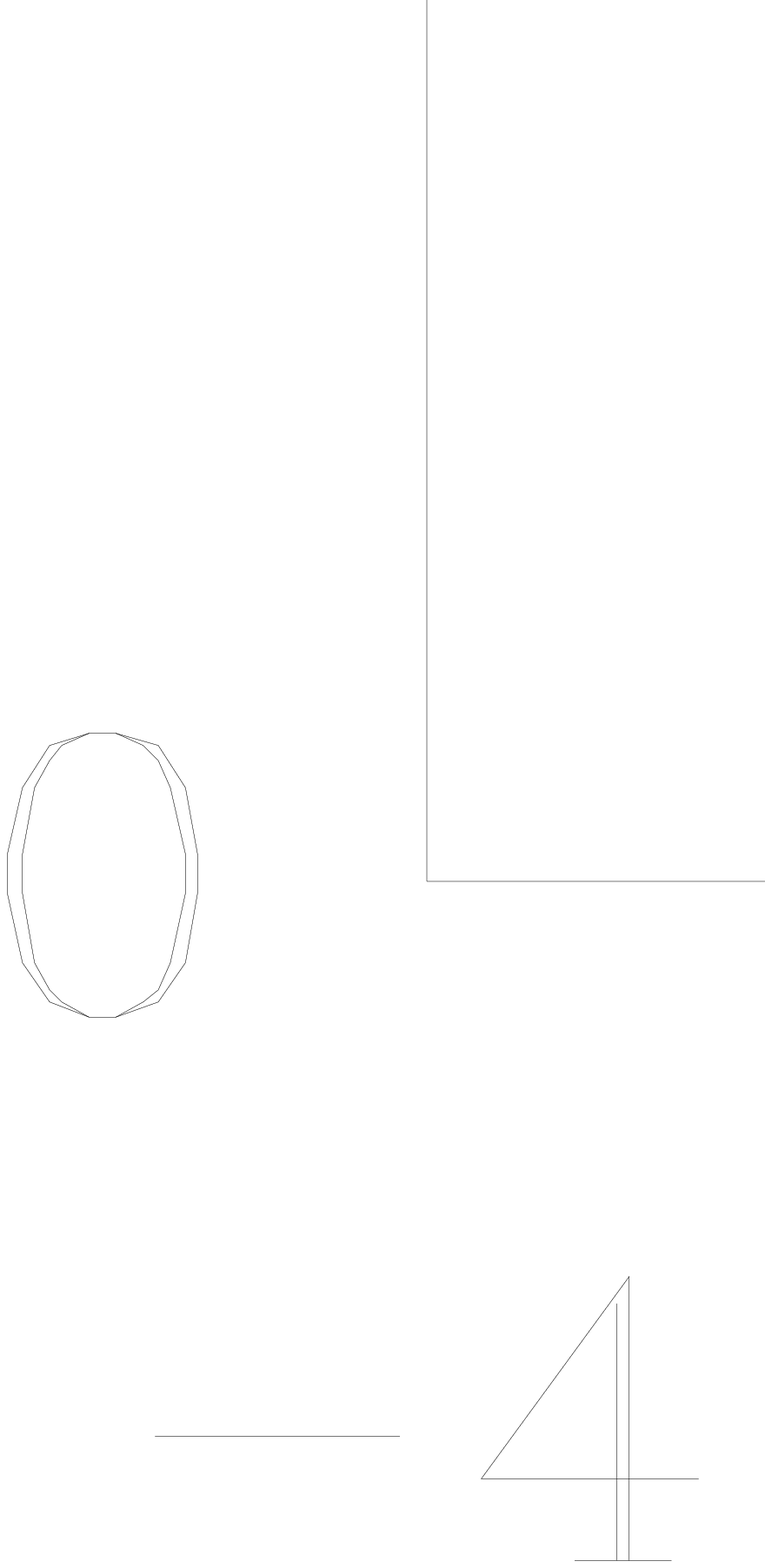,height=5in}}
\indent\vbox{\noindent \hsize 5in Figure 3: A plot of the band 
structure for a periodic set of defects embedded
in a 
material governed by the dispersion relation in (7.7).  For
the defects we have again 
taken $\wd^2 a^2 = 2$, $4\pi N_d\wd^2\Delta^2 a^3 = 1$.
For the parameters of the medium, we have taken
$\wm a = 6$ and 
$32\pi N_m\rho_m\dm^2\wm^2 a^4 = 100$.  The gaps that are too small to
be seen have been explicitly marked out.}
\vskip .2in

The gap between $\wg_+ (k)$ and $\wg_- (k)$ forces the bands
corresponding to these two branches to be separated by this same gap.  
The bands of $\wg_- (k)$ are found below $\wm$ while those of $\wg_+ (k)$
are bounded below by $\sqrt{\wm^2 + 8\pi N_m\rho_m\dm^2}$.  The lower set of
bands away from $\wm$ are similar to those found in Section 6.  For small
$k$ and $\wg$, the spectrum is linear; there is a large gap between the first
and second bands; and there are
smaller gaps between the higher order bands.  The
similarity results from $\wg_- (k)$ being photonic for small $k$.  Close
to $\wm$ the bands of $\wg_- (k)$ accumulate.  They are of an infinite
number, each separated by vanishingly small energies.  The existence of
such bands owes itself to the asymptotic form of $\wg_- (k)$: as
$k\rightarrow \infty$, $\wg_-$ approaches a constant.  The upper set of
bands are photonic away from $\wm$ (provided $\wm > \wd$) 
as $w_+ (k)$ is photonic for large
energies.  Only near $\wm$ do the bands of $\wg_+ (k)$ deviate from
linearity, a result of $\wg_+ (0) \neq 0$.

The gap between the first and second band of $\wg_- (k)$ is 
described as in (6.27):
\eqn\eVIxv{
\Delta E = \bigg| {4\pi N_d \md^2 \over \sqrt{\epsilon (\wd )} 
\sin (\sqrt{\epsilon (\wd )} \wd a)} \bigg| ,}
where $\epsilon (\wd )$ is given by (7.11).
The gaps opening up at $p = n\pi$ for both sets of bands are given
by 
\eqn\eVIxvi{\eqalign{
\Delta_n E_\pm & = |\gamma_\pm (n) | \del_k\omega_\pm (n\pi /a); \cr
\gamma_\pm &= {8 N_d \Delta_d^2 \omega_\pm^2 (n\pi /a) \over
n (\omega_\pm^2(n\pi /a) -\wd^2 )};\cr
\omega_+ (k) = k + & O(\Delta_m^2 ) ;~~~
\omega_- (k) = \wm + O(\Delta_m^2) ; \cr
\del_k\omega_+ (k) = 1 + & O(\Delta_m^2 ) ;~~~
\del_k\omega_- (k) = {8\pi\Delta_m^2 N_m \rho_m \wm k \over 
(k^2 - \wm^2 )^2}.}}
We have again only kept terms to lowest order in
$O(\Delta_d^2)$ and $O(\Delta_m^2)$.
These gaps, as with their counterparts in Section 6, are not symmetric
about $\wg_\pm(n\pi)$.  Rather the gap is defined by 
$\wg = \wg_\pm (n\pi /a)$ and 
$\wg = \wg_\pm (n\pi /a) + {\rm sgn}(\wg_\pm (n\pi /a )^2 -\wd^2)
\Delta_n E_\pm$.


\listrefs
\bye